\documentclass[runningheads]{llncs}

\usepackage{graphicx}
\usepackage{amssymb}
\usepackage{eufrak}
\usepackage{subfig}
\usepackage{booktabs} 
\usepackage{graphicx}
\usepackage{epsfig}
\usepackage{booktabs} 
\usepackage{threeparttable}
\usepackage{colortbl} 
\usepackage{color}
\usepackage{textcomp}
\usepackage{soul} 
\usepackage{comment}
\usepackage{multirow}

\usepackage{amsmath,amsthm}
\theoremstyle{definition}
\newtheorem{Definition}{Definition}

\DeclareMathOperator*{\argmax}{argmax} 
\usepackage{algorithm} 
\usepackage{algpseudocode} 
\begin{document}
\title{Improving Community Detection Performance in Heterogeneous Music Network by Learning Edge-type Usefulness Distribution}
%
%
\author{Zheng Gao\inst{1} \and
Chun Guo\inst{2} \and
Shutian Ma\inst{3} \and Xiaozhong Liu\inst{4}}
\institute{Indiana University Bloomington \and Pandora Media LLC \and Tencent Holdings Ltd \and  Worcester Polytechnic Institute \\
\email{gao27@indiana.edu, cguo@pandora.com, mashutian0608@hotmail.com, xliu14@wpi.edu}}
\maketitle              
\begin{abstract}
With music becoming an essential part of daily life, there is an urgent need to develop recommendation systems to assist people targeting better  songs with fewer efforts. As the interactions between users and songs naturally construct a complex network, community detection approaches can be applied to reveal users' potential interests on songs by grouping relevant users \& songs to the same community. However, as the types of interaction could be heterogeneous, it challenges conventional community detection methods designed originally for homogeneous networks. Although there are existing works on heterogeneous community detection, they are mostly task-driven approaches and not feasible for specific music recommendation. In this paper, we propose a genetic based approach to learn an edge-type usefulness distribution (ETUD) for all edge-types in heterogeneous music networks. ETUD can be regarded as a linear function to project all edges to the same latent space and make them comparable. Therefore a heterogeneous network can be converted to a homogeneous one where those conventional methods are eligible to use. We validate the proposed model on a  heterogeneous music network constructed from an online music streaming service. Results show that for conventional methods, ETUD can help to detect communities significantly improving music recommendation accuracy while simultaneously reducing user searching cost.

\end{abstract}

\section{Introduction}
\begin{figure}[!htb]\centering
	\includegraphics[width=0.8\columnwidth]{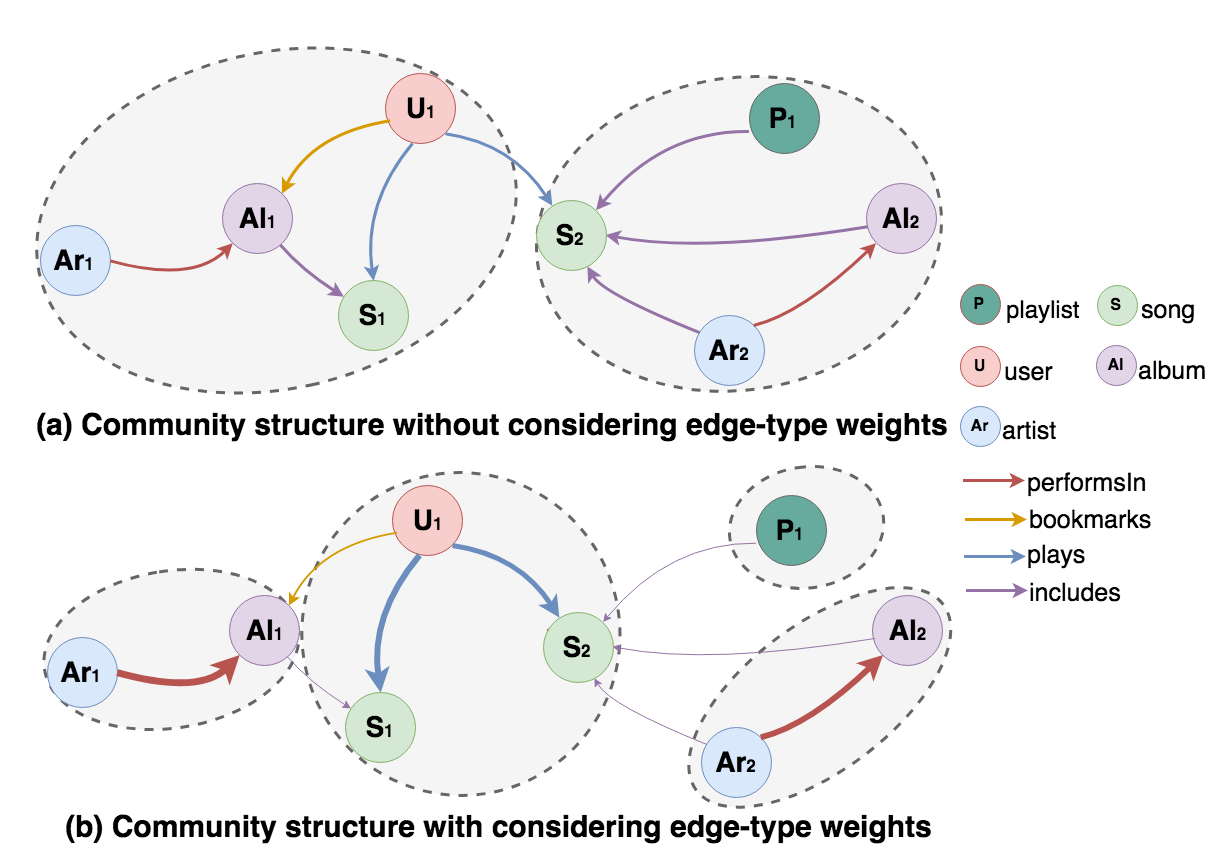}  
	\caption{An example of how ETUD influences the community structure in a heterogeneous music network }
	\label{fig:sample}
	 \vspace{-2em} 
 \end{figure}

 \vspace{-1em}
According to a new report released by Nielsen Music, on average, Americans now spend more than 32 hours a week listening to music\footnote{https://www.forbes.com/sites/hughmcintyre/2017/11/09/americans-are-spending-more-time-listening-to-music-than-ever-before}. Besides, with the boom of online music streaming services (MSS) in the recent years,   user behaviors on MSS (e.g. Pandora and Spotify) become various as well. How to conduct all types of behavior information from MSS to support music recommendation becomes a challenging task. As user behaviors on songs can naturally form a complex heterogeneous network (in Figure \ref{fig:sample}), community detection approaches can be a potential solution to solve this recommendation task by grouping relevant users and songs to the same community. Hence, developing more comprehensive and robust community detection models is in urgent need. In this paper, conventional methods refer to methods are originally designed for homogeneous community detection. Although some other works have addressed on heterogeneous networks, they are mostly task-driven approaches and not able to directly apply on musicrecommendation. In this study, in order to build up a bridge for conventional methods to be eligible on heterogeneous networks, we propose a genetic based method to learn the \textit{edge-type usefulness distribution} (ETUD) on heterogeneous networks in an evolutionary manner. Our model can highlight those edge-types  more important for music recommendation by assigning higher weights on them. After that, all edges are updated by multiplying related ETUD values on their original weights, which converts a heterogeneous network to a homogeneous one.
\vspace{-2em}

Figure \ref{fig:sample} shows an example of how ETUD supports music recommendation  on a sample music network from community viewpoint. Figure ~\ref{fig:sample}(a) ignores the edge-types and treats the network as a homogeneous one. By leveraging a conventional community detection method (Louvain method \cite{blondel2008fast}), the whole music network is partitioned into two communities. while Figure \ref{fig:sample}(b) takes ETUD into account. The thickness of an  edge-type refer to its learned ETUD weight. Obviously the \emph{plays} edge-type is more important for music recommendation. After all edge weights are updated by multiplying related ETUD values, four communities are detected by leveraging the same community detection method. Comparing the two community detection results, it is clear to see that involving ETUD can be beneficial to group the user and his/her favored songs to the same community, which supports music recommendation to achieve better accuracy. 


The contribution of this paper is threefold:

\begin{itemize}
    \item First, a genetic based approach is used to learn ETUD on the heterogeneous networks in an evolutionary manner. It converts a heterogeneous network to a homogeneous one to enable conventional community detection methods on heterogeneous networks as well.
    \item Second, two evaluation tasks are proposed to validate the positive influence of ETUD on heterogeneous networks from searching cost and recommendation accuracy viewpoints. 
    \item  Third, extensive experiments applied on a music online streaming service dataset, Xiami Music, validate that our proposed model is feasible for music recommendation in real cases. 
\end{itemize}


\section{Related works}
\textbf{Homogeneous Community Detection} Current studies follow four main trends \cite{gao2020community}: random walk, embedding, Modularity and overlapping community. node2vec \cite{grover2016node2vec}, DeepWalk \cite{perozzi2014deepwalk} and LINE \cite{tang2015line} are three representative models to learn node embeddings by maximizing the probability to reproduce the network structures. Louvain method \cite{blondel2008fast} designs an agglomerative framework to detect communities with largest Modularity score in an efficient manner.  Infomap \cite{bohlin2014community} assumes a random walker wanders on the network, and minimizes the cost to track the walker's path. Bigclam \cite{yang2013overlapping} formulates overlapping community detection problems into non-negative matrix factorization. \cite{gao2020detecting,ma2017document} leverage cross-domain information for sparse network community detection.

\textbf{Heterogeneous Community Detection} 
\cite{tang2012community} formally describe the heterogeneous community detection problem with a framework of four integration schemes. \cite{huang2018overlapping} aims at overlapping community detection on heterogeneous networks. \cite{comar2012simultaneous} decomposes the heterogeneous network to multiple simple networks. \cite{xie2014community} constructs user profiles from folksonomy systems and use the profiles as auxiliary information for user community detection. \cite{sun2013pathselclus} proposes to use metapath to control node communities with distinct semantics.

\textbf{Community-based Music Recommendation}
\cite{donaldson2007hybrid} considers both user behaviors on playlists and music content profile. \cite{bu2010music} constructs a hypergraph to model the multi-type objects in a music social community. \cite{guo2016feature} focuss on edge-type selection in the heterogeneous network to retain main semantic meanings and trim the network to a smaller scope. Derived from \cite{gao2017personalized}, \cite{gao2019efficient} proposes a generic algorithm for music recommendation guided by user information need.
\section{Method}
\begin{figure}[!htb]\centering
	\includegraphics[width=0.8\columnwidth]{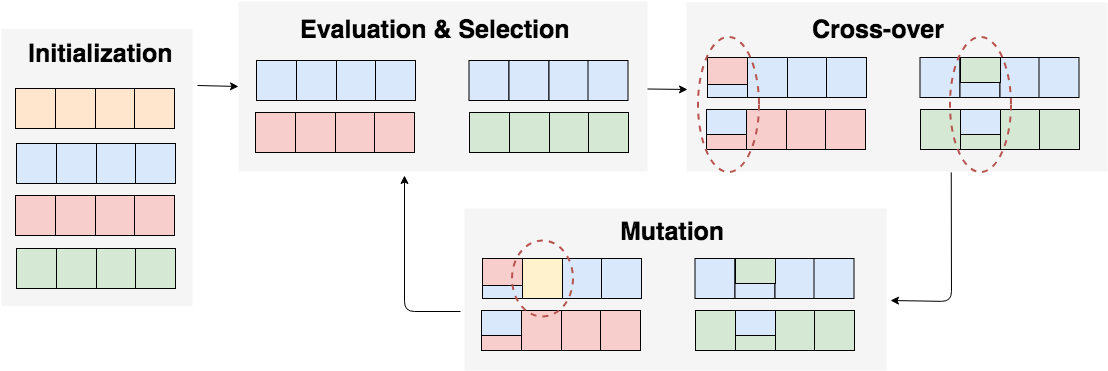}  
	\caption{Genetic approach to learn ETUD in an evolutionary manner}
	\label{fig:pipeline}
	 \vspace{-3em} 
 \end{figure}

 \vspace{-1em}
In this section, a genetic approach is proposed here to learn the \textit{edge-type usefulness distribution} (ETUD). ETUD is the weights for all edge-types to represent their usefulness towards the network. After each edge's weight is updated by multiplying related ETUD score to its original value according to its edge-type, a heterogeneous network is converted to a homogeneous one. Figure \ref{fig:pipeline} shows the proposed method in detail.  

\vspace{-1em} 

\subsection{ETUD Initialization}  
In our genetic approach, the whole process simulates biological evolution process. We define a \emph{chromosome} to represent a possible ETUD result. A \emph{chromosome} is constructed by a set of \emph{gene}s ,where each \emph{gene} refers to an edge-type respectively. The value stores in each \emph{gene} represents the usefulness score of a particular edge-type. A \emph{chromosome} is initialized by randomly assigning value $\in [0,1]$ to each of its \emph{gene} in the beginning. In total, $P$ \emph{chromosome}s are randomly initialized for further evolution process.


\subsection{ETUD Assessment and Selection} 
Given $P$ \emph{chromosome}s, a fitness function   needs to be set up as a criteria to evaluate each \emph{chromosome}'s performance for music recommendation. The best ETUD should be able to best reproduce users' listening preference. In other words, we should find an ETUD that maximizes the similarity between songs' ranking list generated under this ETUD  and user real preference on songs, which is described in Formula \ref{eq:objective}:


\begin{equation}
  \argmax_\textit{w} \sum_{u \in U} sim (g(S|u),p(S|u;w))
  \label{eq:objective}
\end{equation}

$w$ is the  \emph{chromosome} we aim to retrieve, $U$ is the set of all users and $S$ is the set of all songs in the music network.  $g(S|u)$ is the real listening preference of user $u \in U$; $p(S|u;w)$ is all songs ranking list we estimated for user $u$ given the ETUD $w$. Function $sim(\cdot)$ measures the similarity between the two ranking lists. It is hard to leverage stochastic gradient decent on Formula \ref{eq:objective} directly as the differential of $p(S|u;w)$ may be too complex to compute. Genetic approach offers an evolution process to solve this challenge instead. As there are in total $P$ \emph{chromosome}s representing possible ETUD results, we can evaluate each \emph{chromosome} and select better ones to bring offspring \emph{chromosome}s. We call a cycle of selecting \emph{chromosome}s and breeding new offspring \emph{chromosome}s as a \emph{generation}. The whole evolution process will keep running until the fittest  \emph{chromosome} in each \emph{generation} no longer changes. Derived from Formula \ref{eq:objective}, In this paper, each \emph{chromosome}'s fitness value can be calculated as:
\begin{equation}
\textit{$f(w)$} = \frac{exp(\sum_{u \in U} sim( g(S|u),p(S|u;w))}{\sum_{p \in P}exp(\sum_{u \in U} sim( g(S|u),p(S|u;p))} 
\label{eq:fitness}
\end{equation} 

Personalized PageRank algorithm \cite{kloumann2017block} is utilized to represent user $u$ listening preference $p(S|u;w)$. Based on empirical study and common sense that users may only pay attention to the top recommended songs, NDCG@10 \cite{wang2013theoretical} is chosen as the similarity judging function $sim(\cdot,\cdot)$ in this paper. NDCG@10 is an evaluation metric to judge how well the estimated ranking list $p(S|u;w)$ matches the user $u$'s real listening preference $g(S|u)$ and is more sensitive to top ranked results. To represent the fitness score of \emph{chromosome} $w$, we calculate the similarity scores among all users and normalize it via a Softmax function on all \emph{chromosome}s.

In genetic approach, \emph{chromosome}s with higher fitness score should have higher chance to be preserved for the next-round evolution. Hence, we use bootstrap sampling strategy (random sampling with replicates) to randomly sample $P$ \emph{chromosome}s based on their fitness scores. This sampling strategy has two advantages: First, there are always $P$ \emph{chromosome}s in each generation, which ensures a stable evolution process; Second, \emph{chromosome}s with higher fitness scores are more likely to be preserved, so that the best ETUD result is more likely to be found after running a number of \emph{generation}s. 

\subsection{ETUD Transformation} 

The initialized \& selected \emph{chromosome}s can not fully cover the whole possible ETUD results. Hence, there is a need to search the whole ETUD result space and find out the best one. To achieve this in an efficient manner, two transformation strategies including  Cross-over and Mutation are applied. These two strategies help to generate new \emph{chromosome}s from the the original selected \emph{chromosome}s, which achieves the \emph{chromosome} evolution. 

Specifically, In Cross-over step, all selected \emph{chromosome}s are grouped into pairs first, and the Cross-over transformation will be applied within pairs. Given two \emph{chromosome}s in the same pair, which are denoted as $w_1$ and $w_2$.  In this step, $w_1$ and $w_2$ exchange values stored in partial of their genes to create two offspring \emph{chromosome}s. $w_{1k}$ refers to the edge-type usefulness weight stored in the $k_{th}$ gene of $w_1$. The whole process is showed as: 
\begin{equation}
w_{ij}^{*} = 
(\frac{w_{ij}}{ w_{1j}+ w_{2j}})^{p_j}w_{ij}^{(1-p_j)}
\label{eq:transform}
\end{equation}
\begin{eqnarray}\text{$p_j$}=
\begin{cases}
\text{$1$}, & r_j \ge t_c\cr 
\text{$0$}, &  r_j < t_c
\end{cases}
\label{eq:binary}
\end{eqnarray}
where $i \in \{1,2\}$ refers to the index of the two newly generated \emph{chromosome}s. $w_{ij}$ is the value of the $j_{th}$ gene in $w_{i}$. $p_j$ is an binary indicator to decide whether the cross-over step occurs on the $j_{th}$ gene or not. The value of $p_j$ is controlled by a randomly generated variable $r_j \in [0,1]$ and a pre-defined threshold $t_c \in [0,1]$. 

Unlike the Cross-over step to generate new offspring \emph{chromosome}s via the interactions between two original \emph{chromosome}s, Mutation step allows a \emph{chromosome} to generate a new offspring \emph{chromosome} within itself. For \emph{chromosome} $w_{1}$, instead of exchanging values of genes with another \emph{chromosome} $w_{2}$, it exchanges its genes' value with a random variable $\mathcal{X}$ randomly drew from normal distribution $\mathcal{X} \sim \mathcal{N}(0,1)$. And the gene values of new \emph{chromosome} is still calculated via Formula \ref{eq:transform} and \ref{eq:binary} with another pre-defined threshold $t_m \in [0,1]$.


\subsection{ETUD Finalization}

The whole evoluntion process will keep running for \emph{generation}s until the best \emph{chromosome} no longer changes for $\mathbb{N}$ generations. And the \emph{chromosome} with largest fitness value in the last \emph{generation} will be the returned ETUD.

However, there is still an edge-type dependency issue remained. One example viewed in Figure \ref{fig:sample} is that edge-type \emph{performIn} and edge-type \emph{plays} are independent so that their weights are not comparable. To address this, we define: 

\begin{Definition}
Two edge-types are dependent only if they share either same start-node type or same end-node type. And only weights of dependent edge-types are comparable.
\label{def:dependency}
\end{Definition}

Following Definition \ref{def:dependency}, all edge-types are grouped into several independent sets first ( In Figure \ref{fig:sample}, the edge-type sets are \{\emph{performsIn}, \emph{includes}\} and \{\emph{plays}, \emph{bookmarks}\}). After that, ETUD weights of edge-types in the same set are normalized via the formula:
\begin{equation}
\textit{$ew_i^{*}$} = \frac{ew_{i}}{\sum_{S(e_{i}) = S(e_{j})}ew_{j}} 
\end{equation} 

$e_{i}$ refers to the $i_{th}$ edge-type; $ew_{i}$ refers to the returned ETUD weight of edge-type $e_{i}$; $S(e_{i})$ is the edge-type set which $e_{i}$ belongs to. And the normalized $ew_i^{*}$ is regarded as the final edge-type usefulness weight of $e_{i}$. 
\section{Experiment}
 \vspace{-1em}
\subsection{Dataset Description \& Parameter Setting}
 \begin{table}
\centering
    \renewcommand{\tabcolsep}{3pt}
    \begin{tabular}{|l|l|l|l|l|l|} \hline
    \textbf{Node-type}   &\textbf{Count} & \textbf{Node-type}   &\textbf{Count}  &\textbf{Node-type}&\textbf{Count}       \\ \hline 
    
    Song& 54,353& User & 38,780& Genre& 543\\ \hline 
   Playlist &47,098 & Artist & 9,901& Album& 17,730\\ \hline 
 \end{tabular}
         \caption{Node statistics}
         \label{tab:node}
 \end{table} 
 
    
 
 \begin{table}
        \centering
      \vspace{-1em} 
    \renewcommand{\tabcolsep}{2pt}
    \begin{tabular}{|l|l|l|l|l|l|} \hline
    \textbf{Edge}   &\textbf{Name} & \textbf{Count}   & \textbf{Edge}   &\textbf{Name} & \textbf{Count}        \\ \hline 
    
    user$\to$song &plays &3,991,226 & user$\to$playlist&makes &42,775  \\ \hline 
    playlist$\to$song& includes& 514,652 &album$\to$song &include &40,798 \\ \hline 
    user$\to$artist&plays & 238,862& user$\to$artist&comment & 36,813\\ \hline 
    user$\to$song &bookmarks &213,602 &artist$\to$album &performsIn & 17,457\\ \hline
    user$\to$song&comments &136,078 & album$\to$genre&categorizedAs &15,174 \\ \hline
    user$\to$album&bookmarks &89,342 &user$\to$playlist &bookmark &9,982 \\ \hline 
    user$\to$album&comments & 59,575& artist$\to$genre& categorizedAs &9,726 \\ \hline 
    artist$\to$song&performsIn &57,675 &user$\to$playlist &comment &9443 \\ \hline 
 \end{tabular}
         \caption{Edge statistics}
         \label{tab:edge}
 \vspace{-2em} 
 \end{table} 

In this paper, the dataset used for music recommendation is from Xiami, one of the largest online music streaming services in China. Based on user behaviors and song profiles, a complex heterogeneous network is constructed. Details of the network are showed in Table \ref{tab:node} and Table \ref{tab:edge}. Besides this network, we also have all users listening history record as ground truth. For each user, all his/her listened songs are labelled as a score $\in \{1,2,3,4\}$, representing the quartile of their play counts belongs to. The listening history of 70 percent users are randomly sampled used to training our model ( used in Section 3.2), and the rest 30 percent users are used for our model validation.

The parameters used in our approach are set based on empirical experiments and previous studies : number of initialized \emph{chromosome} $P$ = 1000; Cross-over threshold $t_c$ = 0.95 and Mutation threshold $t_m$ = 0.1; termination criteria is when the best \emph{chromosome} no longer changes for $\mathbb{N} = 10$ generations.






\subsection{Searching Cost with Between-community Jumping} 

The first task is to examine whether applying ETUD on heterogeneous networks can reduce the searching cost of retrieving all users' listened songs.  

In our music network, each user $u$ will belong to a community after a community detection method is leveraged. When user $u$ looks for songs to listen, there is a searching cost for the user to retrieve all his/her favored songs. We simulate user searching behaviours and define the overall searching cost as:

\begin{equation}
    Cost(U_{t}, S) = \sum_{u \in U_{t}}\sum_{k = 1}^{K}\sum_{s \in C_{u}^{k}}\frac{\sum_{s \in C_{u}^{k}}\mathbb{I}(s|u)}{\varphi(s|u)\sum_{s \in S}\mathbb{I}(s|u) }\cdot log(\lVert C_{u}^{(k-1)}\rVert \cdot \lVert C_{u}^{k}\rVert)  
    \label{eq:cost}
\end{equation}

$U_{t}$ is the collection of all testing users and $S$ are all songs in the heterogeneous network. There are $K$ communities generated by leveraging conventional methods. As in each community, the number of songs that $u$ listened before can be calculated, all communities are ranked by their contained number of listened songs in the descending order. Therefore $C_{u}^{k}$ means the community containing the $k_{th}$ most number of user $u$'s listened songs. $\rVert C_{u}^{k}\lVert$ is the number of nodes in this community. To make it easier to calculate, we define $\rVert C_{u}^{0}\lVert = 1$. $\varphi(s|u)$ is the number of times that user $u$ listens song $s$. $\mathbb{I}(s|u)$ is binary indicator $\in \{0,1\}$ to judge whether user $u$ listened the song before or not. 

This searching cost function depicts how much effort users take to retrieve all their listened songs by jumping between communities. For each user $u$, the  community ranking reveals how user $u$ favors to each community. The user $u$ starts to search from the most favoured community to the least favoured community in a descending sequence until all listened songs are retrieved. Formula \ref{eq:cost} defines a searching cost for all testing users jumping between communities. Community sizes of previous community and current community are considered during community jumping as larger community will take users more effort to find the listened songs. The searching cost also has a reciprocal relationship with the number of times users listened on songs because for more favored songs, the searching cost to retrieve them should be less. 
We calculate communities generated with/without ETUD with six classic conventional community detection algorithms (See Section 2 for algorithm description) and compare the searching cost of retrieving the top 5,10,20,50,100 listened songs for all testing users (songs are ranked by the number of listening times). The result is shown in Table \ref{tab:cost}.

  \begin{table}
        \centering
    \renewcommand{\tabcolsep}{1pt}
    \begin{tabular}{|l|l|l|l|l|l|l|l|l|} \hline
       \textbf{Algorithm} &  \textbf{Category} & \textbf{Edge-type} & \textbf{ Comm. \#}   &\textbf{Cost@5} &\textbf{Cost@10}  &\textbf{Cost@20}   &\textbf{Cost@50}   &\textbf{Cost@100}  \\ \hline 
         \multirow{2}{*}{DeepWalk \cite{perozzi2014deepwalk}}  &  \multirow{2}{*}{embedding}    &  No  &100    &  197.29   &    420.01 & 897.37  &2309.50  &4450.39\\ 
           &      &  Yes   &  100 & \textbf{187.99}    &\textbf{402.51}    &\textbf{856.04}   &\textbf{2054.70}  &\textbf{4121.64} \\ \hline 
         \multirow{2}{*}{LINE \cite{tang2015line}}  &  \multirow{2}{*}{embedding}    &  No  &100    &   172.03 &   379.08  & 811.87  & 2106.40 & 4134.17\\ 
           &      &  Yes   & 100  &  \textbf{162.29}  & \textbf{371.29}    & \textbf{783.70}  & \textbf{2012.18} & \textbf{3922.18}\\ \hline 
         \multirow{2}{*}{node2vec \cite{grover2016node2vec}}  &  \multirow{2}{*}{embedding}    &  No  &   100 & 181.84  &  391.83 &849.79  & 2175.11 & 4132.87\\ 
           &      &  Yes   & 100  & \textbf{173.60}  & \textbf{390.29}    & \textbf{824.02}  & \textbf{2096.56} & \textbf{3984.89}\\ \hline 
        \multirow{2}{*}{Louvain \cite{blondel2008fast}}  &  \multirow{2}{*}{modularity}    &  No  & 12   & 577.40   &1275.62 & 2695.33  & 6925.05 & 12845.07\\ 
           &      &  Yes   & 38  & \textbf{533.19}   &  \textbf{1092.83} &  \textbf{2350.49} & \textbf{5963.32} &\textbf{10532.66} \\ \hline 
         \multirow{2}{*}{Infomap \cite{bohlin2014community}}  &  \multirow{2}{*}{random walk}    &  No  &  3705  &  850.09    & 1918.62   &  4022.01 & 10372.20 &19784.88 \\ 
           &      &  Yes   & 4507  &\textbf{682.42}   & \textbf{1571.91}  & \textbf{3358.09}  &\textbf{8494.20} &\textbf{15997.72} \\ \hline 
         \multirow{2}{*}{BigClam \cite{yang2013overlapping}}  &  \multirow{2}{*}{overlapping}    &  No  &   100 & 11.23 &    27.80 & 56.73  & 131.87& 270.42\\ 
           &      &  Yes   & 100  & \textbf{7.91}   &  \textbf{16.68}  & \textbf{39.36} & \textbf{105.74} & \textbf{183.10}\\ \hline 
       
         \end{tabular}
         \caption{Searching cost results comparison}
         \label{tab:cost}
         \vspace{-2em}
 \end{table} 
 Although different algorithms have different-scale searching costs and community number, all community results learned with ETUD require less searching cost than the communities without considering ETUD significantly. It means that learning ETUD via our model on heterogeneous network can reduce users' effort to search their favoured songs.

\subsection{Searching Accuracy within Community}

After exploring how well ETUD can help to reduce searching costs, we are also willing to see whether it also benefits music recommendation accuracy within communities. Ideally, the learned ETUD should have positive effect to group users and their favoured songs into the same community. In this task, for each testing user $u$, we select all songs in the same community as the user, and rank the songs based on their PageRank scores calculated during the training process. NDCG is still the evaluation metric used to evaluate how well the generated PageRank result matches user real listening history. The averaged NDCG score for all testing users is in Table \ref{tab:ranking}.
From the result, under all circumstances, NDCG scores with ETUD are significantly higher than scores without ETUD. It infers that by taking ETUD into account, it is more likely to group users and their favoured songs into the same community. 

  \begin{table}
    \centering
    \renewcommand{\tabcolsep}{2pt}
    
    \begin{tabular}{|c|c|c|c|c|c|} \hline
       \textbf{Algorithm} &  \textbf{Edge-type}    &\textbf{NDCG @5} &\textbf{NDCG@10}  &\textbf{NDCG@20}   &\textbf{NDCG@100}   \\ \hline 
         \multirow{2}{*}{DeepWalk}  &    No    & 0.6932  &  0.6974   &  0.6971  &  0.5741   \\ 
           &    Yes   &  \textbf{0.7529}  &  \textbf{0.7355} & \textbf{0.7162} & \textbf{0.5992}    \\ \hline 
         \multirow{2}{*}{LINE}  &     No &   0.4883 &0.3588    &  0.2530  &  0.1055   \\ 
           &   Yes   &\textbf{0.5153} & \textbf{0.4141}  & \textbf{0.3057} &  \textbf{0.1401} \\ \hline 
         \multirow{2}{*}{node2vec}  &   No   & 0.6995   &  0.7007 & 0.6967  & 0.5796  \\ 
           &  Yes    &  \textbf{0.7476}  & \textbf{0.7349}  &  \textbf{0.7206} &  \textbf{0.5924} \\ \hline 
        \multirow{2}{*}{Louvain}  &   No  &  0.6764  & 0.6742   &  0.6829  &  0.7168   \\ 
           &    Yes  & \textbf{0.7370}  & \textbf{0.7366} &  \textbf{0.7328} & \textbf{0.7328}    \\ \hline 
         \multirow{2}{*}{Infomap}  &    No   & 0.7020 & 0.6955 & 0.6974 & 0.7264  \\ 
           &   Yes   & \textbf{0.7548}  & \textbf{0.7430}  & \textbf{0.7332}  &  \textbf{0.6927}  \\ \hline 
         \multirow{2}{*}{BigClam}  &   No &  0.5963  &   0.5193 &  0.4524  &  0.2626  \\ 
           &  Yes    &  \textbf{0.6479} & \textbf{0.5793}  & \textbf{0.5022}  &  \textbf{0.2376}   \\ \hline 
       
         \end{tabular}
         \caption{Within-community music retrieval accuracy evaluation}
         \label{tab:ranking}
         \vspace{-4em}
 \end{table}

\section{Conclusion}
 \vspace{-1em}
In this paper, we put efforts on how to convert a heterogeneous network to a homogeneous one so that conventional community detection methods can still be eligible to use. It extends the application scope of those conventional methods and endows them more generalizable and robust usage. To achieve this, a genetic based approach is developed to learn the edge-type usefulness distribution (ETUD) on heterogeneous networks. Experiments on a real music dataset  show that involving ETUD in heterogeneous community detection is able to facilitate the accuracy of music recommendation with less search cost significantly. In the future, we will explore more sophisticated algorithms such as reinforcement learning methods for ETUD estimation. 

\bibliographystyle{splncs04}
\bibliography{acmart} 
\end{document}